\documentstyle[12pt,epsfig,wrapfig]{article}
\newcommand{\be}{\begin{equation}}
\newcommand{\ee}{\end{equation}}
\newcommand{\bea}{\begin{eqnarray}}
\newcommand{\eea}{\end{eqnarray}}
\newcommand{\la}{\langle}
\newcommand{\ra}{\rangle}
\begin{document}
\title{On the azimuthal asymmetries in DIS}
\author{A.V.Efremov$^a$, K.Goeke$^b$,
M.V.Polyakov$^{b,c}$, D.Urbano$^{b,d}$\\
\footnotesize\it $^a$ Joint Institute for Nuclear Research, Dubna, 141980
Russia\\
\footnotesize\it $^b$ Institute for Theoretical Physics II, Ruhr University
Bochum, Germany\\
\footnotesize\it $^c$ Petersburg Nuclear Physics Institute,
Gatchina, 188350 Russia \\
\footnotesize\it $^d$ Faculdade de Engenharia da Universidade do Porto, 4000
 Porto, Portugal}
\date{}
\maketitle

\vspace{-9cm}
\begin{flushright}
RUB/TP2-02/00
\end{flushright}
\vspace{7cm}

\begin{abstract}
Using the recent experimental data on the left right asymmetry in
fragmentation of transversely polarized quarks and the theoretical
calculation of the proton transversity distribution in the effective
chiral quark soliton model we explain the azimuthal asymmetries
in semi-inclusive hadron production on longitudinally (HERMES) and
transversely (SMC) polarized targets with no free parameters.
On this basis we state that the proton transversity distribution
could be successfully measured in future DIS experiments with
{\it longitudinally} polarized target.
\end{abstract}
%%%%%%%%%%%%%%%%%%%%%%%%%%%%%%%%%%%%%%%%%%%%%%
%%%%%%%%%%%%%%%%%%%%%%%%%%%%%%%%%%%%%%%%%%%%%%

\section{Introduction}

Recently a first information about azimuthal asymmetries in
semi inclusive hadron production on longitudinally (HERMES
\cite{hermes}) and transversely (SMC  \cite{bravardis99}) polarized
targets was reported.  These asymmetries contain information on the proton
transversity distribution.

The three most important (twist-2) parton distributions
functions (PDF) in a nucleon are the non-polarized distribution function
$f_1(x)$, the longitudinal spin distribution  $g_1(x)$ and the transverse
spin distribution  $h_1(x)$ \cite{transversity}. The first two have
been more or less successfully measured experimentally in classical
deep inelastic scattering (DIS) experiments but the measurement of
the last one is especially difficult since it belongs to the class of
the so-called helicity odd structure functions and can not be seen
there. To access  these helicity odd structures one needs either to
scatter two polarized protons or to know the transverse polarization
of the quark scattered from transversely polarized target.

There are several ways to do this:
\begin{enumerate}
\item
To measure a polarization of a self-analyzing hadron to which the
quark fragments in a semi inclusive DIS (SIDIS), e.g.
$\Lambda$-hyperon \cite{augustin}.  The drawback of this method
however is a rather low rate of quark fragmentation into
$\Lambda$-particle ($\approx 2\%$) and especially that it is mostly
sensitive to $s$-quark polarization.
\item
To use a spin dependent T-odd parton fragmentation function (PFF)
\cite{muldt,muldz,mulddis} responsible for the left-right asymmetry
in one particle fragmentation of transversely polarized quark with
respect to quark momentum--spin plane. (The so-called "Collins
asymmetry" \cite{collins}.)
\item
To measure a transverse handedness in multi-particle parton
fragmentation \cite{hand}, i.e. the correlation of the quark spin
4-vector $s_\mu$ and particle momenta $k^\nu$,
$\epsilon_{\mu\nu\sigma\rho}s^\mu k_1^\nu k_2^\sigma k^\rho$
($k=k_1+k_2+k_3+\cdots$).  \end{enumerate}

The last two methods are comparatively new and only in the last years
some experimental indications to the T-odd PFF have appeared
\cite{todd,czjp99}. In this paper we will use this result to extract
the information on the proton transversity distribution.  More
exactly with this result and the calculation of $h_1(x)$ in the
effective chiral quark soliton model \cite{pp96}, we compute the
azimuthal asymmetries in SIDIS and compare it with the experimental
points \cite{hermes,bravardis99}.
A similar work was done recently in the paper \cite{aram}
where the authors used some adjustable parametrization for the
T-odd PFF and some estimations for  $h_1(x)$. Our approach is free
of any adjustable parameters.

In \cite{DPPPW96} it was shown that the chiral quark--soliton model
possesses all features needed for a successful description of the
nucleon parton structure: it is essentially a quantum
field-theoretical relativistic model with explicit quark degrees of
freedom, which allows an unambiguous identification of the quark as
well as the antiquark distributions in the nucleon. Owing to its
field-theoretical nature the quark and antiquark distributions
obtained in this model satisfy all general QCD requirements:
positivity, sum rules, inequalities, etc.

Analogous of $f_1,\ g_1$ and $h_1$ are the functions $D_1,\ G_1$ and
$H_1$, which describe the fragmentation of a non-polarized quark into
a non-polarized hadron  and a longitudinally or transversely
polarized quark into a longitudinally or transversely polarized
hadron, respectively\footnote{We use the notation of the work
\cite{muldt,muldz,mulddis}.}.

These fragmentation functions are integrated over the transverse 
momentum $\mathbf{k}_T$ of a quark with respect to a hadron. With 
$\mathbf{k}_T$ taken into account, new possibilities arise. Using the 
Lorentz- and P-invariance one can write in the leading twist 
approximation 8 independent spin structures \cite{muldt,muldz}. Most 
spectacularly it is seen in the helicity basis where one can build 8 
twist-2 combinations, linear in spin matrices of the quark and hadron 
{\boldmath$\sigma$}, $\mathbf{S}$ with momenta $\mathbf{k}$, 
$\mathbf{P}_h$.

Especially interesting is a new T-odd and helicity odd  structure that
describes a left--right asymmetry in the fragmentation of a transversely
polarized quark:
%\newline
$
H_1^\perp\mbox{\boldmath$\sigma$}({\mathbf k}\times
{\mathbf P}_{h\perp})/k\la P_{h\perp}\ra,
$
where the coefficient $H_1^\perp$ is a function of the longitudinal
momentum fraction $z$ and quark transverse momentum  $k_T^2$. The
$\la P_{h\perp}\ra$ is the averaged transverse momentum of the final
hadron\footnote{ Notice different normalization factor compared to
\cite{muldt,muldz,mulddis}, $\la P_{h\perp}\ra$ instead of $M_h$.}.

Since the $H_1^\perp$ term is helicity odd, it makes possible to
measure the proton transversity distribution $h_1$ in semi-inclusive
DIS from a transversely polarized target by measuring the left-right
asymmetry of forward produced pions (see~\cite{mulddis,kotz} and
references therein).

The problem is that this function was completely unknown both
theoretically and experimentally.  Meanwhile, the data collected by
DELPHI (and other LEP experiments) give a possibility to measure
$H_1^\perp$.  The point is that despite the fact that the transverse
polarization of a quark ( an antiquark) in Z$^0$ decay is very small
($O(m_q/M_Z)$), there is a non-trivial correlation between transverse
spins of a quark and an antiquark in the Standard Model:  $C^{q\bar
q}_{TT}= {(v_q^2-a_q^2)/(v_q^2+a_q^2)}$, which reaches rather high
values at $Z^0$ peak: $C_{TT}^{u,c}\approx -0.74$ and
$C_{TT}^{d,s,b}\approx -0.35$.  With the production cross section
ratio $\sigma_u/\sigma_d=0.78$ this gives for the average over
flavors a value of $\la C_{TT}\ra\approx -0.5$.

The spin correlation results in a peculiar azimuthal angle dependence
of produced hadrons, if the T-odd fragmentation function $H_1^\perp$
does exist~\cite{collins,colpsu}.  A simpler method has been proposed
recently by an Amsterdam group \cite{muldz}. They predict a specific
azimuthal asymmetry of a hadron in a jet around the axis in direction
of the second hadron in the opposite jet\footnote{ We assume the
factorized Gaussian form of $P_{h\perp}$ dependence for 
$H_1^{q\perp}$ and $D_1^q$ integrated over $|P_{h\perp}|$.}:
\begin{eqnarray}
{{\rm d}\sigma\over {\rm d}\cos\theta_2 {\rm d}\phi_1}\propto
(1+\cos^2\theta_2)\cdot \left(1+ {6\over\pi}\left[{H_1^{\perp q}\over
D_1^q}\right]^2 C_{TT}^{q\bar q}{\sin^2\theta_2\over
  1+\cos^2\theta_2}\cos(2\phi_1)\right) \, ,
\label{mulders}
\end{eqnarray}
where $\theta_2$ is the polar angle of the electron and the second 
hadron momenta $\mathbf{P}_2$, and $\phi_1$ is the azimuthal angle 
counted off the $(\mathbf{P}_2,\, \mathbf{e}^-)$-plane.
\begin{wrapfigure}{HR}{6.5cm}
%\raisebox{-10mm}{
\mbox{\epsfig{figure=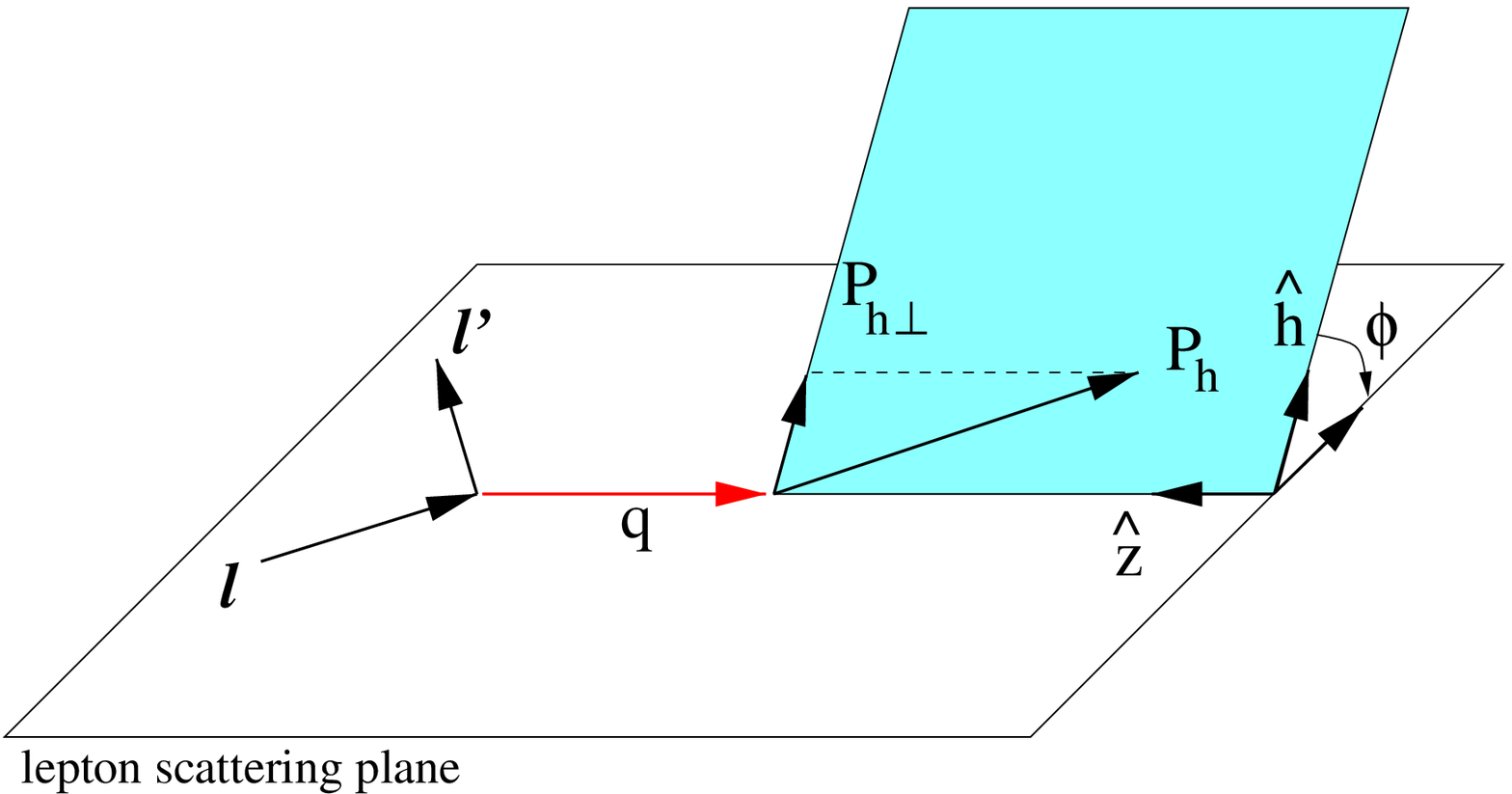,width=6.5cm,height=6.0cm}}%}\\

\bigskip
{\footnotesize\it
{\bf Figure 1.}~
Kinematics of SIDIS in the Lab frame.
}
\end{wrapfigure}

This asymmetry was measured \cite{todd} using the DELPHI data
collection.  For the leading particles in each jet of two-jet events,
summed over $z$ and averaged over quark flavors (assuming
$H_1^{\perp}=\sum_H H_1^{\perp\, q/H}$ is flavor independent), the
 most reliable value of the analyzing power is given by
\begin{equation}
\left|{\la H_1^{\perp}\ra\over\la D_1\ra}\right|
=(6.3\pm 2.0)\% \, ,
\label{apower}
\end{equation}
with presumably large systematic errors\footnote{Close value
was also obtained from pion asymmetry in inclusive $pp$-scattering
\cite{bogl99}.}.

\section{Azimuthal asymmetries}

The T-odd azimuthal asymmetry in semi inclusive DIS $ep\to
e'\pi^{\pm}X$, which was measured by HERMES consist of two sorts of
terms (see \cite{mulddis} Eq.  (115)): a twist-2 asymmetry
$\sin2\phi_h$ and a twist-3 asymmetry $\sin\phi_h$. Here $\phi_h$
is the azimuthal angle around the $z$-axis opposite to direction of
virtual $\gamma$ momentum in the Lab frame, counted from the electron
scattering plane (see Fig. 1).  The first asymmetry is proportional
to the $p_T$-dependent transverse quark spin distribution in
longitudinally polarized proton, $h_{1L}^\perp(x,p_T)$, while the
second one contains two parts:  one term is proportional to the
twist-3 distribution function  $h_L(x)$ and the second one
proportional to the twist-3 interaction dependent correction to the
fragmentation function $\widetilde H_L$. In what follows we will
systematically disregard this interaction dependent correction such
as correction $\widetilde H_L$ or $\widetilde h_L$ \footnote{The
calculations in the instanton model of QCD vacuum supports this
assumption \cite{DPG99}.}.

In the same approximation, the integrated functions over the quark
transverse momentum, $h_{1L}^\perp(x,p_T)$ and $h_L(x)$, are
expressed through $h_1$ (see \cite{mulddis} Eqs. (C15),(C19))
\be
h_{1L}^{\perp(1)}(x)\equiv\int d^2p_T\left(\frac{p_T^2}{2M^2}\right)
h_{1L}^{\perp}(x,p_T)=-x^2\int_x^1d\xi h_1(\xi)/\xi^2=-(x/2)h_L(x) \, ,
\label{wwform}
\ee
and the longitudinal spin dependent part of the SIDIS integrated over
the modulus of the final hadron transverse momentum (assuming a Gaussian
distribution) and over $z$ reads as\footnote{Subscript $U,L,T$ means
unpolarized, longitudinally or transversely polarized beam or target
respectively.} (see \cite{mulddis} Eq. (115))

\bea
\nonumber
\frac {d\sigma_{UL}}{dxdyd\phi}&=&\frac{2\alpha^2s}{Q^4}S_L
\Biggl[\frac{2M}{\la P_{\perp h}\ra}\frac{1-y}{1+
\frac{\la p_T^2\ra}{\la k_T^2\ra}}\sin2\phi
\cdot x^3\sum_a e_a^2\left(\int_x^1d\xi h^a_1(\xi)/\xi^2\right)
\la H^{\perp a/\pi}_1(z)\ra\\
&+&\frac{M}{Q}\frac{4(2-y)\sqrt{1-y}}{\sqrt{1+
\frac{\la p_T^2\ra}{\la k_T^2\ra}}}
\sin\phi\cdot x^3\sum_a e_a^2\left(\int_x^1d\xi h^a_1(\xi)/\xi^2\right)
\la H^{\perp a/\pi}_1(z)/z\ra
\Biggr] \, ,
\label{sigl}
\eea
where $M$ is the nucleon mass, $S_L=S\cos\theta_\gamma\approx
S(1-2M^2x(1-y)/sy)\approx S$ is longitudinal with respect to the
virtual photon\footnote{Notice, that in HERMES experiment the target
polarization was longitudinal with respect to the incident electron
beam.} part of the proton polarization $S$, $s=2(Pl)=2ME$ is the
electron-proton c.m. total energy squared, $Q^2=sxy$ the modulus of
the squared momentum transfer $q=l-l'$, $x=Q^2/2(Pq)$, $y=2(Pq)/s$.
The quantities $\la p_T^2\ra$ and  $\la k_T^2\ra=\la P^2_{h\perp}\ra/z^2$ 
are mean square values of the transverse momenta of the quark in 
distribution and fragmentation functions, respectively.

For the transverse part of the polarization and for the unpolarized 
target one can write (see \cite{mulddis} Eq. (110) with $\phi_S=\pi$ 
and Eq.  (113))
\bea
\frac {d\sigma_{UT}}{dxdyd\phi}&=&\frac{2\alpha^2s}{Q^4}S_T
\frac{1-y}{\sqrt{1+\frac{\la p_T^2\ra}{\la k_T^2\ra}}}\sin\phi
\cdot x\sum_a e_a^2 h^a_1(x)\la H^{\perp a/\pi}_1(z)/z\ra \, ,
\label{sigt}  \\
\frac {d\sigma_{UU}}{dxdyd\phi}&=&\frac{2\alpha^2s}{Q^4}
\cdot\frac{1+(1-y)^2}{2}
\cdot x\sum_a e_a^2 f^a_1(x)\la D_1^{a/\pi}(z)\ra \, ,
\label{sigun}
\eea
where $S_T=S\sin\theta_\gamma\approx S\sqrt{4M^2x(1-y)/sy}$ is the
transverse part of the proton polarization.

The asymmetries measured by HERMES are
\be
A_{UL}^W=\frac{\int d\phi dyW \left(d\sigma^+/S^+dxdyd\phi
-d\sigma^-/S^-dxdyd\phi\right)}
{\frac{1}{2}\int d\phi dy\left(d\sigma^+/dxdyd\phi
+d\sigma^-/dxdyd\phi\right)}\ ,
\label{asymw}
\ee
where $W=\sin\phi$ or $\sin2\phi$ and $S^\pm_H$ is the nucleon
polarization\footnote{More preferable would be the weights
$W=(P_{h\perp}/\la P_{h\perp}\ra)\sin\phi$ or
$(P_{h\perp}^2/M\la P_{h\perp}\ra)\sin2\phi$ since the factors
$(1+\la p_T^2\ra/\la k_T^2\ra)$ in denominators of (\ref{sigl}) and 
(\ref{sigt}) would disappear.} 
($\pm$ sign means different spin directions with "+" means opposite 
to incident lepton beam), averaged over the transverse momentum 
$P_{h\perp}$ and over $z$ of the final $\pi^+$ or $\pi^-$ and 
$\sigma=\sigma_{UU}+\sigma_{UL}+\sigma_{UT}$.

Substituting (\ref{sigun}), (\ref{sigl}), (\ref{sigt}) into 
(\ref{asymw}) and integrating over the region of $y$ allowed by the 
experimental cuts:
\be
1\,GeV^2\le Q^2\le 15\,GeV^2,\ W^2=(P+q)^2\ge 4\,GeV^2,\ y<0.85 \, ,
\label{cuts}
\ee
one can find the asymmetries $A_{UL}^{\sin\phi}$ and $A_{UL}^{\sin2\phi}$
proportional to the ratios
\be
\frac{\sum_a e_a^2 h^a_1(x)\la H^{\perp a/\pi}_1(z)/z\ra}
{\sum_a e_a^2 f_1^a(x)\la D_1^{a/\pi}(z)\ra} \, ,
\label{atrans}
\ee
and
\be
\frac{x^2\sum_a e_a^2\left(\int_x^1d\xi h^a_1(\xi)/\xi^2\right)
\la H^{\perp a/\pi}_1(z)\ra}
{\sum_a e_a^2 f_1^a(x)\la D_1^{a/\pi}(z)\ra} \, .
\label{along}
\ee

\begin{wrapfigure}{HR}{5.7cm}
%\raisebox{-5mm}{
\mbox{\epsfig{figure=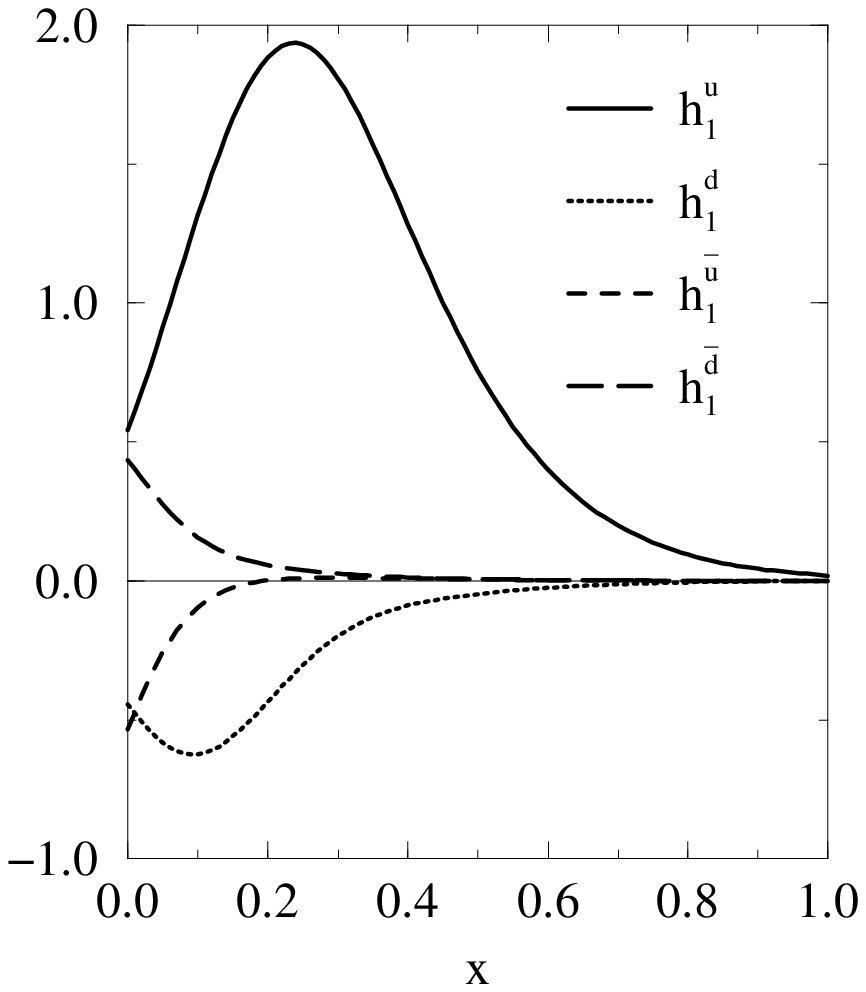,width=6.0cm,height=6cm}}\\%}
{\footnotesize\it
{\bf Figure 2.}~Transversity quark and antiquark distributions at a low
normalization point ($\mu\sim 600MeV$) in the effective chiral quark
soliton model.}
\end{wrapfigure}

Let us assume that only the favored fragmentation functions 
$D_1^{a/\pi}$ and $H_1^{\perp a/\pi}$ will contribute this ratios, i.e.
$
D_1^{u/\pi^+}(z)=D_1^{\bar d/\pi^+}(z)=D_1^{d/\pi^-}(z)
=D_1^{\bar u/\pi^-}(z)\equiv D_1(z)
\nonumber
$
and similarly for $H^{\perp}_1(z)$. The dominance of the favored 
T-odd fragmentation is asserted also from Sch\"afer-Teryaev sum rule 
for these functions \cite{ST99}.  To explore the DELPHI result 
(\ref{apower}) we will use the approximation $\la 
H^{\perp}_1(z)/z\ra= \la H^{\perp}_1(z)\ra/\la z\ra$ with the 
experimental values $\la z\ra=0.41$ and $\la P_{h\perp}\ra\approx \la 
p_T\ra\approx 0.4\,GeV$ (see \cite{mrst98} and \cite{delphipt}).  
This would allow us to extract from the observed HERMES asymmetries 
an information on $h_1^u(x)+(1/4)h_1^{\bar d}(x)$ and to compare with 
some model prediction.  Instead, we use the prediction of the chiral 
soliton model for $h_1^a(x)$ (see Fig. 2) and the GRV parametrization 
\cite{GRV} of the unpolarized DIS data for  $f_1^a(x)$ to calculate 
the asymmetries $A_{UL}^{\sin\phi}$ and $A_{UL}^{\sin2\phi}$ for 
$\pi^+$ and $\pi^-$.  The comparison of the asymmetries thus obtained 
with the  HERMES experimental data is presented on Fig.3.

%
%\vspace{-4cm}
\begin{figure}[ht]
\begin{center}
\raisebox{-40mm}{
\mbox{\epsfig{figure=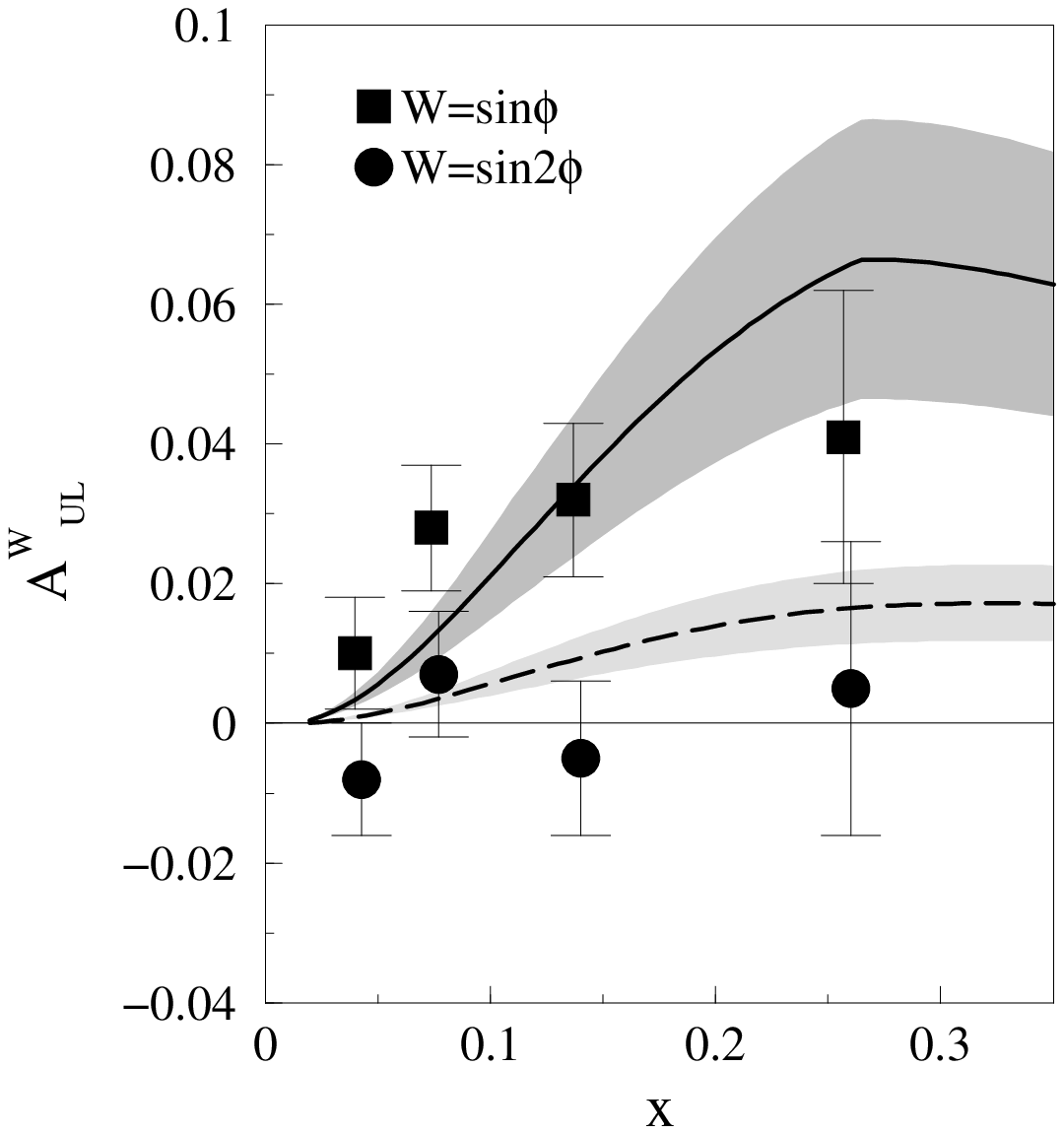,width=6.5cm,height=6.5cm}}}
\raisebox{-40mm}{
\mbox{\epsfig{figure=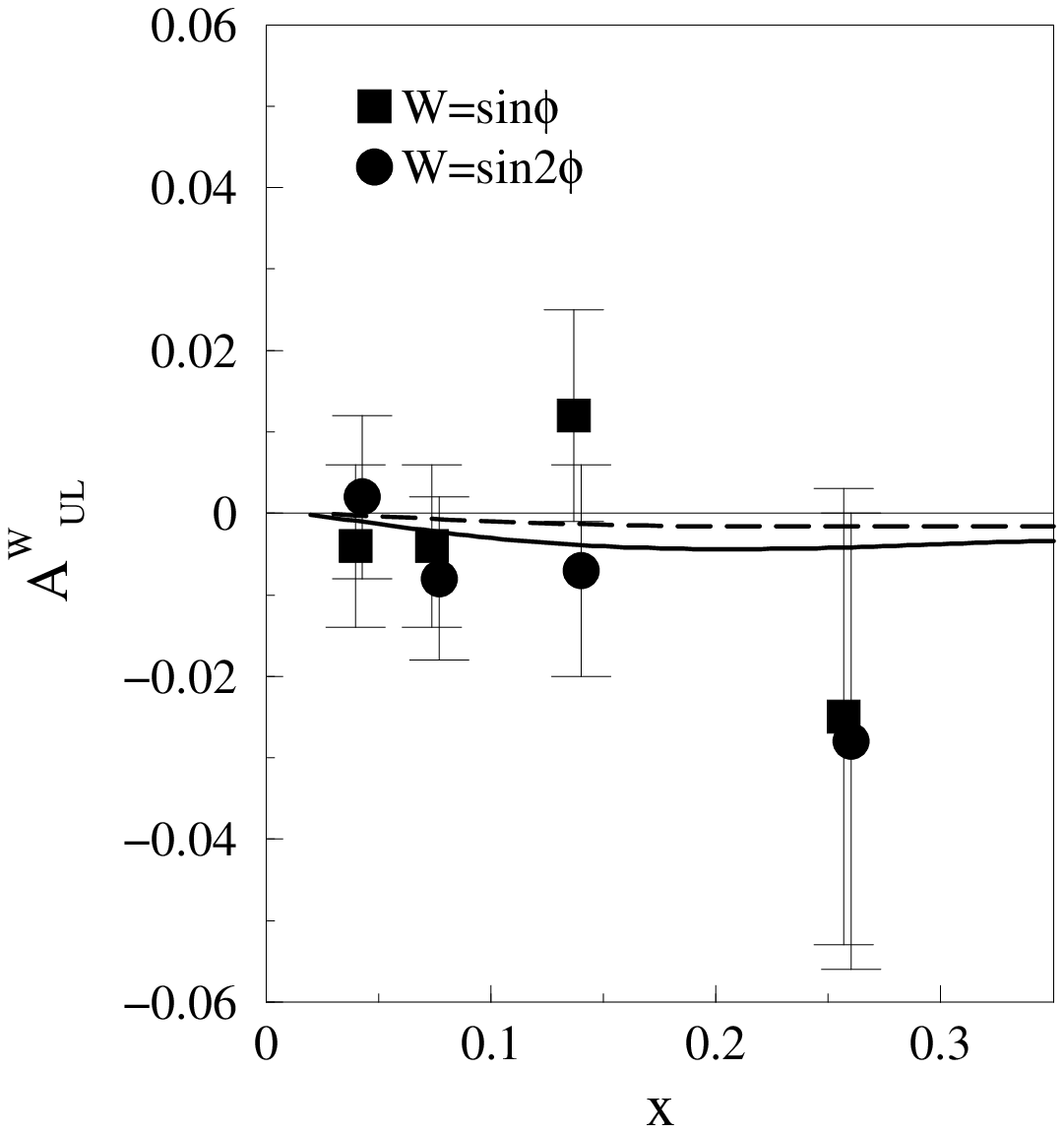,width=6.5cm,height=6.5cm}}}\\
\end{center}
{\footnotesize\it
{\bf Figure 3.} Single spin azimuthal asymmetry for
$\pi^{+}$ ({\bf left}) and $\pi^{-}$ ({\bf right}):
$A_{UL}^{\sin\phi}$ (squares) and $A_{UL}^{\sin2\phi}$ (circles) as a
functions of x. The solid ($A_{UL}^{\sin\phi}$) and the dashed lines
($A_{UL}^{\sin2\phi}$) correspond to the chiral quark-soliton
model calculation at
$Q^2=4\,GeV^2$. The shaded areas on the left figure represent the
experimental uncertainty in the value of the ratio
$\left|{\la H^{\perp}_1 (z)\ra}/{\la D_1(z)\ra}\right|$.
For $\pi^-$ (right) both asymmetries are compatible with zero.}
\end{figure}

The agreement is good enough though the experimental errors are yet 
rather large. Moreover the sign of the asymmetry is uncertain since 
only the modulus of the analyzing power (\ref{apower}) is known 
experimentally.  However, Fig.3 gives evidence for positive sign.  
The theoretical curves correspond to normalization point 
$Q^2=4\,GeV^2$ although the $Q^2$-dependence of the asymmetries is 
very weak and do not exceeds 10\% in the range (\ref{cuts}).  Notice 
that in spite of the factor ${M}/{Q}$ the $\sin\phi$ term in Exp.  
(\ref{sigl}) is several times larger than that of $\sin2\phi$ for 
moderate $Q^2$. That is why this asymmetry prevails for the HERMES 
data where  $\la Q^2\ra\approx 2.5\,GeV^2$.  One can thus state that 
the effective chiral quark soliton model \cite{pp96} gives a rather 
realistic picture of the proton transversity $h_1^a(x)$.

The interesting observable related to $h_1(x)$ is the proton tensor 
charge defined as
\begin{equation}
g_T\equiv \sum_a\int_0^1 dx\Big(h_1^a(x)-h_1^{\bar a}(x)\Big) \, .
\end{equation}
The  calculation of $g_T$ in this model  yields  for
$Q^2=4\,GeV^2$ \cite{pol96}
\be
g_T=0.6 \, .
\ee
The most recent experimental value of the proton axial charge is
\cite{leader99}
\be
a_0=0.28\pm0.05 \, ,
\ee
and the value obtained in the model is $a_0 = 0.35$ 
\cite{anomaly,BMG}.  We can conclude that the chiral quark soliton 
model predicts very different values for the axial and tensor charges 
of the nucleon, which is in contradiction with the nonrelativistic 
quark model prediction.

Concerning the asymmetry observed by SMC \cite{bravardis99} on
transversely  polarized target one can state that it agrees with
result of HERMES.  Really, SMC has observed the azimuthal asymmetry
${\rm d}\sigma(\phi_c)\propto const\cdot(1+a\sin\phi_c)$, where
$\phi_c=\phi_h+\phi_S-\pi$ ($\phi_S$ is the azimuthal angle of the
polarization vector) is the so-called Collins angle. The raw
asymmetry $a=P_T\cdot f\cdot D_{NN}\cdot A_N$, where $P_T,\, f,$ and
$D_{NN}=2(1-y)/\left[1+(1-y)^2\right]$ are the target  polarization
value, the dilution factor and the spin transfer coefficient.

The physical asymmetry $A_N$, averaged over transverse momenta 
(assuming again a Gaussian form) is given by expression 
(\ref{atrans}) divided by $\sqrt{1+\la p_T^2\ra /\la k_T^2\ra }$.  
With the same functions $h_1^a(x),\ f_1^a(x)$, integrating over $x$ 
separately the numerator and the denominator weighted by $xQ^{-4}$, 
we get for $Q^2=4\,GeV^2$ with error due to (\ref{apower}) (see Fig. 4)
\be
A_N=0.07\pm0.02\, ,
%0.108
\ee
which should be compared with the experimental best fit value
$A_N=0.11\pm0.06$.

%
%\vspace{-4cm}
\begin{figure}[ht]
\begin{center}
%\raisebox{-40mm}{
\mbox{\epsfig{figure=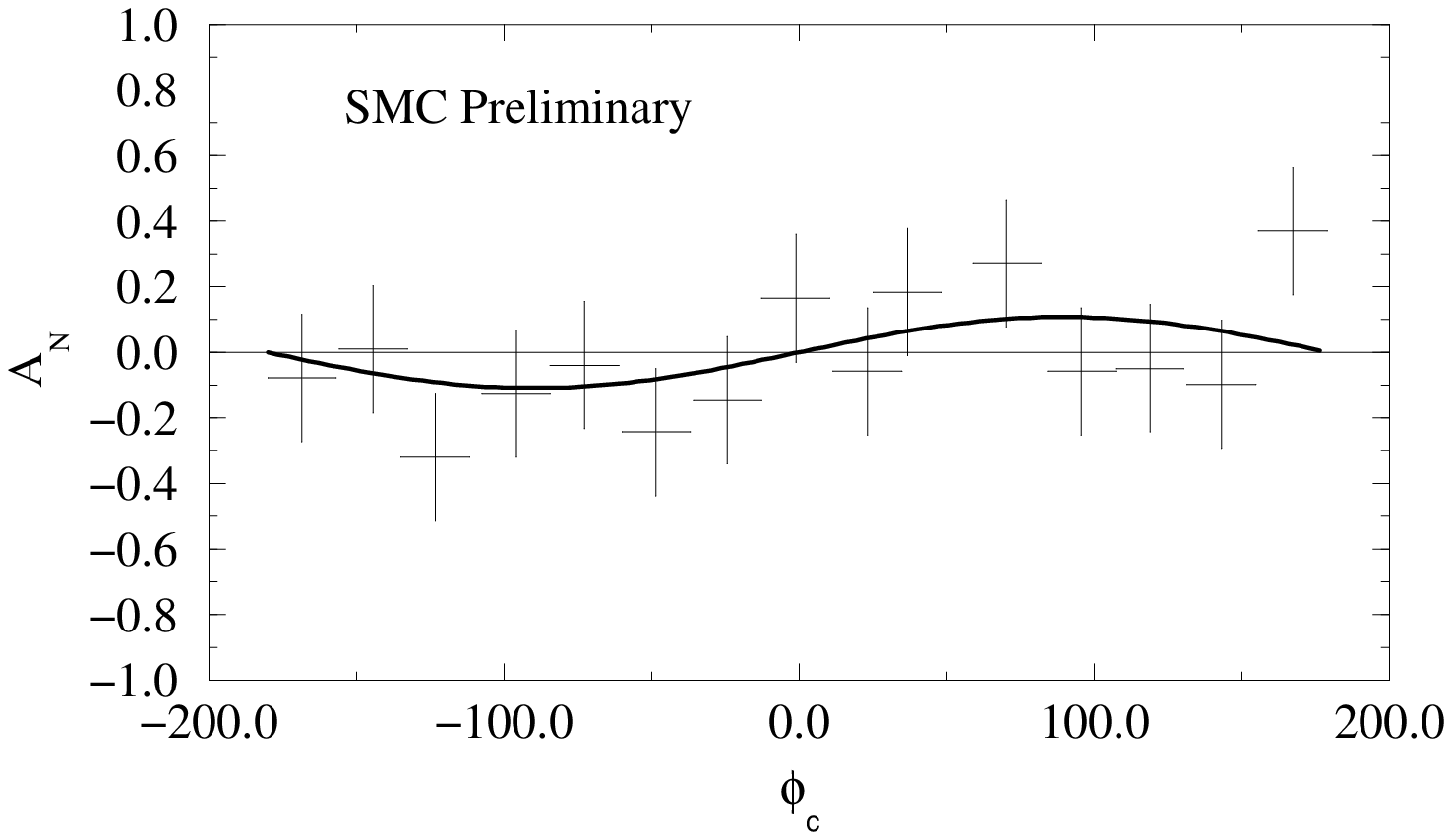,width=10cm,height=6.0cm}}%}
\end{center}
{\footnotesize\it
{\bf Figure 4.}
Collins angle $\phi_c$ azimuthal distributions for positive hadrons
produced off transversely polarized protons.  The solid line is the
curve  $0.07\sin\phi_c$, obtained in the
the chiral quark-soliton model.}
\end{figure}

\section{Conclusions}
In conclusion, using the  effective chiral quark soliton model
for the proton transversity distribution  we obtain a rather good
description of the azimuthal asymmetries in semi-inclusive hadron
production measured by HERMES and SMC, though the experimental errors
 are yet large. This, however is only the first experiment! We would
like to stress that  our description has no free adjustable
parameters.

Probably the most useful lesson we have learned is that to measure
transversity in SIDIS in the region of moderate $Q^2$ it is not
necessary to use a transversely polarized target. Due to
approximate Wandzura-Wilczek type relations (\ref{wwform}) one can
explore the longitudinally polarized target also.  This is very
important for future experiments, like COMPASS at CERN since the
proton transversity measurements could be done simultaneously with
measurement of the spin gluon distribution $\Delta g(x)$. For a better
interpretation of the result a better knowledge of quark analyzing
power (\ref{apower}) with smaller systematic errors is necessary.

\medskip
{\small We would like to thank H.~Avakian, D.~Boer, A.~Kotzinian,
P.~Mulders, A.~Sch\"afer, O.~Teryaev, P.~Pobylitsa for fruitful
discussions. One of authors (A.E.) is grateful to the Institute for
Theoretical Physics II of Ruhr University Bochum, where main part of
this work was done, for warm hospitality.
D.U. acknowledges financial
support from PRAXIS XXI/BD/9300/96 and from
PRAXIS PCEX/C/FIS/6/96. The work has partially been supported by the
DFG and BMBW}

%%%%%%%%%

\end{document}